\documentclass[pra,twocolumn]{revtex4}
\usepackage{graphicx}
\usepackage{dcolumn}
\usepackage{bm}
\usepackage{amsmath}
\usepackage{amsfonts}

\newcommand{\bra}[1]{\langle #1|}
\newcommand{\ket}[1]{|#1\rangle}

\begin{document}
\title{Observing the evolution of a quantum system that does not evolve }
\author{Simone \surname{De Liberato}}
\affiliation{Laboratoire Mat\'eriaux et Ph\'enom\`enes
Quantiques, Universit\'e Paris 7, \\ B\^atiment Condorcet, 10 rue
Alice Domont et L\'eonie Duquet, 75205 Paris Cedex 13, France}
\affiliation{Laboratoire Pierre Aigrain, \'Ecole Normale
Sup\'erieure, 24 rue Lhomond, 75005 Paris, France}

\date{\today}

\begin{abstract}
This article deals with the problem of gathering information on the time evolution of a single metastable quantum system whose evolution is impeded by the quantum Zeno effect. It has been found it is in principle possible to obtain some information on the time evolution and, depending on the specific system, even to measure its average decay rate, even if the system does not undergo any evolution at all.
\end{abstract}

 \maketitle

\section{Introduction}


A quantum system initially prepared in a state that is not an energy eigenstate will generally undergo an evolution. For very short evolution times, the transition probabilities towards other states increase as the square of the time, and a metastable quantum system, which is observed frequently enough, would remain essentially unchanged. This is the quantum Zeno effect, pictorically described as the fact that, in quantum mechanics, "a watched pot never boils". This effect has been subject to intense theoretical  \cite{Zeno,Zeno1,Zeno2,Zeno3,Zeno4,Zeno5} and experimental \cite{ZenoEx1, ZenoEx2} investigations.


The question this article wants to assess is whether  any information about the time evolution, and, above all, the nondiagonal elements of the Hamiltonian that determine the transition rates, can be measured while the system is being forbidden to evolve by a series of projective measurements performed frequently enough to be in the quantum Zeno regime.

At a first glance, the answer may seem to be negative, each measurement projecting the system wavefunction in its initial state and, upon measurement, all the information about time-evolved states of the system being lost.

Moreover, if such procedure existed, it would give us the possibility to perform  some quite paradoxical measurements when applied to normally irreversible dynamics.

Given an unknown quantum system in an excited state, undergoing an irreversible spontaneous decay, we could be able to know if the system is stable or not before the decay takes place. We could even manage to gather information on its average decay rate, not only by measuring a single system instead of a statistical ensemble, but keeping this single system undecayed.


We will show that, given the possibility to make the system interact coherently with a properly crafted environment, such measurement is indeed possible.

The measured quantity will be a certain function of the coupling with all the possible final states; the actual amount of information we can obtain will depend largely on the specific system and how much we know about it.

This work is organized in the following way: in section II, the general ideas exploited in the rest of the paper are introduced. These ideas will be used in sections III to lay down the full quantum mechanical calculations and the efficency assessment of the measuring procedure. Finally, in section IV, the developed theory will be applied to some specific model Hamiltonians.

\section{General framework}

Let us suppose we have a certain system, whose evolution is being slown down by a series of projective measurements. If the measurements are frequent enough, there is a high probability that each one of them will find the system in its initial state.
In order to gather information on the time evolution of the system, the only thing we can do, in the time interval between two successive projective measurements, is to make it interact with some other measuring apparatus.

The system, found in its initial state by one of the repeated projective measurements, is left to evolve freely for a certain fraction of time before the following projective measurement; then we make it interact coherently  with some ancillary quantum system, such that the state of the system is recorded in the ancilla. Let us say that a $0$ will be recorded if the system is found in its initial state and a $1$ if it is in some other state. Then the system is left to evolve again until the next projective measurement, that still finds it in the initial state. Now there are no correlations anymore between the system and the ancilla, we can thus check the ancilla state independently from the state of the system.
The state of the ancilla just after the projective measurement will not be a $0$ but a certain linear superposition of $0$ and $1$, where the coefficient of the $1$ is generally not zero. It is due to the finite probability that the system, during the first part of the evolution (before the interaction with the ancilla), evolves to some other state and, during the second part (after the interaction with the ancilla), evolves back to the initial state.

\begin{figure}[htbp]
\begin{center}
\includegraphics[width=7.5cm]{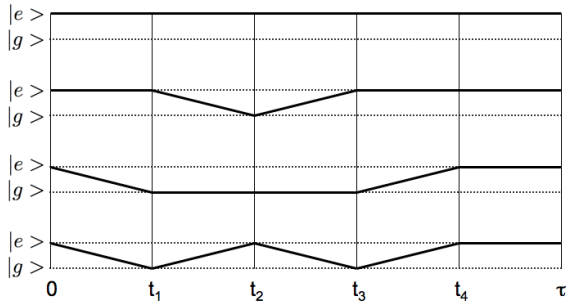}
\caption{\label{highorder} A schematization of higher order processes in the simple case of a two levels system. The system interacts four times with the environment at $t=t_1,t_2,t_3$ and $t_4$. The projective measurements are performed at $t=0$ and $t=\tau$ and both of them find the system in its excited state. The solid lines show the trajectory of the system. The first one corresponds to a zeroth order process, with no jumps, the second and the third to a second order process, and the fourth to a fourth order process. Clearly, if in the end the system is to be found in its initial state, only even order processes are possible. }
\end{center}
\end{figure}

Even if the projective measurements always find the system in its initial state, because of the quasi-reversible dynamics typical of the quantum Zeno regime, we could rely on these sort of higher order processes to gather information about its time evolution.
In order to give a more formal and general definition of a higher order process, let us suppose that, during the time interval between two successive projective measurements, the system interacts several times with an environment, so that each interaction leaves in it a record of the state of the system. We will call a n-th order process a branch in which the system interacts at least $n$ times with the environment and the environment records exactly $n$ jumps, both from and to the initial state (fig. \ref{highorder}).

According to this definition, in the previous example, the nonzero coefficient of the $1$ is due to a second order process.

Obviously, the probability of a n-th order process  will generally be of the order of the n-th power of the probability for the system to jump only one time, that is the probability of it being found by the projective measurement in a state that is not its initial state.
This probability vanishing in the quantum Zeno regime, it would seem impossible to gather any information on a single system while still blocking its evolution. It will be shown that on the contrary, by using a carefully crafted interaction with the environment, instead of the simple one of the previous example, it is indeed possible to make the probability of the second order processes high enough to measure the desired quantity, while still finding the system in its initial state.

\section{Formal theory}

Let  $\ket{e}$ be the initial state of our system and $\ket{g_i}$ a set of mutually orthogonal states, orthogonal to $\ket{e}$, such that $\ket{e}$ and all the $\ket{g_i}$ form an orthonormal base for the Hilbert space of the system.  The index $i$ will in general run over a discrete as well as continuous part. We will consider as ancilla a two levels system (a qubit) prepared in the state, for the moment arbitrary,  $\ket{ \psi}$.
Let $H$ be the Hamiltonian of the unperturbed system and $K(t)=e^{-iHt/\hbar}$ its evolution operator. The composite system is prepared in the state $\ket{\varphi(0)}=\ket{e} \ket{\psi}$. A projective measurement is then performed on the system each $\tau$ that, with a certain probability, projects the system state upon the state $\ket{e}$.
We will suppose $\tau$ to be small enough to allow us to neglect all terms  of order greater than two in the time development of $K(\tau)$.

We let the system evolve freely for a time $\tau/2$, neglecting the auto Hamiltonian of the ancilla.
The resulting state of the composite system will be

\begin{eqnarray*}
\ket{\varphi(\tau^-/2)}&=&K(\tau/2)\ket{e}\ket{ \psi}\\&=&\lbrack K_{ee}\ket{e} +\sum_i K_{ie} \ket{g_i} \rbrack \ket{ \psi},
\end{eqnarray*}

where $K_{ee}=\bra{e}K(\tau/2) \ket{e}$, $K_{ie}=\bra{g_i} K(\tau/2) \ket{e}$ and $K_{ei}=\bra{e}K(\tau/2) \ket{g_i}$.
Then we entangle the ancilla with the system by performing a controlled-M on it, conditionally on the system being or not in its initial state, that is we act on the ancilla with the identity operator if the system is found in its initial state and with the operator $M$ otherwise.
The operator M is for the moment a to be determined unitary operator that acts only on the ancilla, whose two eigenvalues we will name $e^{i\nu_+}$ and $e^{i \nu_-}$and the respective eigenvectors $\ket{+}$ and $\ket{-}$. Let assume the time to perform the controlled-M  to be much smaller than $\tau$, in order to be allowed to ignore the effect of the free evolution. We obtain

\begin{eqnarray*}
\ket{\varphi(\tau^+/2)}=\lbrack K_{ee}\ket{e} +M\sum_i K_{ie}  \ket{g_i}\rbrack \ket{ \psi}.
\end{eqnarray*}

After a second evolution of $\tau/2$, just before the next projective measurement, the state will be

\begin{eqnarray*}
\label{tau-}
\ket{\varphi(\tau^-)}&=&K(\tau/2)\ket{\varphi(\tau^+/2)}\\&=&
\lbrack (K_{ee}^2+M\sum_i K_{ei}K_{ie})\ket{e} \\&+&\sum_i (K_{ie}K_{ee}+M\sum_j K_{ij}K_{je}  )\ket{g_i}\rbrack \ket{ \psi}.
\end{eqnarray*}

The projective measurement will find the system in the state $\ket{e}$ with a probability

\begin{eqnarray}
P_e=\lvert K_{ee}^2\lvert^2 +\lvert\Delta K_{ee}^2\lvert^2+2Re\lbrack \bar{K}_{ee}^{2}\Delta K_{ee}^2\bra{\psi}  M \ket{ \psi}\rbrack,
\end{eqnarray}

where we use the completeness relation $\mathbb{I}=\ket{e}\bra{e} +\sum_i \ket{g_i}\bra{g_i}$ to write

\begin{eqnarray}
\label{variance}
&&\sum_i K_{ei}K_{ie}=\bra{e} K(\tau/2)(\mathbb{I}-\ket{e}\bra{e})K(\tau/2)\ket{e}\nonumber \\
&&=\bra{e}K(\tau/2)^2\ket{e}-\bra{e}K(\tau/2)\ket{e}^2=\Delta K_{ee}^2.
\end{eqnarray}

Supposing we have effectively found the system in its initial state, therefore the state of the composite system will be

\begin{eqnarray}
\label{tau+}
\ket{\varphi(\tau^+)}=
\frac{K_{ee}^2+M\Delta K_{ee}^2}{\sqrt{P_e}}\ket{e} \ket{ \psi}=U_{eff}\ket{e} \ket{ \psi}.
\end{eqnarray}

The operator $U_{eff}$ acts only on the ancilla's Hilbert space, so that we can say that, with a probability $P_e$, the system is left unchanged and the ancilla undergoes an effective evolution $U_{eff}$.

The probability $P_e$ is given by the expression
\begin{eqnarray}
\label{P_e_1}
P_e=\lvert K_{ee}^2\lvert^2 +\lvert\Delta K_{ee}^2\lvert^2+2Re\lbrack \bar{K}_{ee}^{2}\Delta K_{ee}^2\bra{\psi} M \ket{ \psi}\rbrack,
\end{eqnarray}

where it is worthwhile noticing that $P_e$, and consequently  $U_{eff}$, depend on the initial state of the ancilla, due to the fact that the effective evolution operator comes from a non-unitary dynamics.

Since we are interested in the Zeno regime, we can make a development of $K(\tau/2)$ to the second order, obtaining
\begin{eqnarray}
\label{op_develop}
K(\tau/2)=1-\frac{i\tau H}{2\hbar}-\frac{\tau^2 H^2}{8\hbar^2}+O(\frac{\tau^3 H^3}{\hbar^3})
\end{eqnarray}

and we can notice that by shifting the zero energy level in order to have $\bra{e}H \ket{e}=0$, we can make both $H_{ee}$ and $\Delta H_{ee}$ real up to the second order in time. We have thus, inserting eq. \ref{op_develop} into eq. \ref{variance},

\begin{eqnarray}
\label{develop_K}
K_{ee}^2&=&1-\frac{\tau^2 \Delta H^2_{ee}}{4\hbar^2}+O(\tau^3  \bar{\Omega}^3)\\
\Delta K_{ee}^2&=&-\frac{\tau^2 \Delta H^2_{ee}}{4\hbar^2}+O(\tau^3 \bar{\Omega}^3) \nonumber
\end{eqnarray}

where $\Delta H_{ee}=\sqrt{\bra{e}H^2 \ket{e}}$ and $\bar{\Omega}$ is a certain complex number whose norm is not bigger than the norm of the Hamiltonian's biggest matrix element divided $\hbar$.

Inserting the developments of eq. \ref{develop_K}  into  eq. \ref{tau+} and \ref{P_e_1}, we see that, due to the reality of $K_{ee}^2$ and $\Delta K_{ee}^2$, if we take $\nu_+=\pm \nu_-$,  $U_{eff}$ does not alter the relative weight of the two eigenvectors up to the considered order. We will thus choose $\nu_+=- \nu_-=\nu$.

The  eigenvectors of $M$ are also in this case approximate eigenvectors of $U_{eff}$:

\begin{eqnarray}
\label{U_eff_dev}
U_{eff}\ket{\pm}&=&\lbrack1\mp \frac{i\tau^2\Delta H_{ee}^2\sin(\nu)}{4\hbar^2}+O(\tau^3  \bar{\Omega}^3)\rbrack \ket{\pm},
\end{eqnarray}

approximate, because in the error term there is a dependence upon $\ket{\psi}$.

The probability $P_e$ becomes

\begin{eqnarray}
\label{P_e_dev}
P_e&=&1-\frac{\tau^2 \Delta H^2_{ee}(1+\cos(\nu))}{2\hbar^2}+O(\tau^3  \bar{\Omega}^3).
\end{eqnarray}

Taking $\ket{\psi}$ as an equal weighted superposition of the two eigenvectors  of $M$, that is  $\ket{\psi}=\frac{1}{\sqrt{2}}(\ket{+}+ \ket{-})$, we obtain from eq. \ref{tau+} and \ref{U_eff_dev}

\begin{eqnarray*}
\label{tau+_final}
\ket{\varphi(\tau^+)}
&=&\ket{e}\frac{e^{i\phi} \ket{+} + e^{-i\phi} \ket{-}}{\sqrt{2}},
\end{eqnarray*}

where the phases $\phi$ and $\nu$ are linked by the relation:

\begin{eqnarray}
\label{phi}
\phi=-\frac{\tau^2 \Delta H^2_{ee}\sin(\nu)}{4\hbar^2}+O(\tau^3  \bar{\Omega}^3).
\end{eqnarray}

The ratio between $\phi$ (the signal we want to measure) and $1-P_e$ (the transition probability) plays for us the role of a signal to noise ratio (SNR), that is the quantity we have to maximize when we choose the operator M,

\begin{eqnarray}
\label{SNR}
\text{SNR}=\frac{\sin(\nu)+O(\tau  \bar{\Omega})}{2(1+\cos(\nu))+O(\tau  \bar{\Omega})}.
\end{eqnarray}

This quantity is not limited as a function of $\tau$ and $\nu$, being divergent for
$\nu\rightarrow \pi$ and $\tau\rightarrow 0$ with

\begin{eqnarray}
\label{condition}
\lvert \pi-\nu \lvert \gg \sqrt{\tau  \bar{\Omega}}.
\end{eqnarray}

Repeating the same procedure $N$ times on the same ancilla, with $N$ of the order of $1/\phi$,
we obtain

\begin{eqnarray*}
\label{Ntau+}
\ket{\varphi(N\tau^+)}
&=&\ket{e}\frac{e^{iN\phi} \ket{+} + e^{-iN\phi} \ket{-}}{\sqrt{2}}.
\end{eqnarray*}

At this point we can measure the state of the ancilla in the base $\frac{1}{\sqrt{2}}(\ket{+} \pm \ket{-})$, and the probability of the two outcomes will be $\cos(N\phi)^2$ and $\sin(N\phi)^2$.
Being $N\phi$ of the order of the unity, repeating the whole procedure a small number of times will permit us to ascertain it with the desired precision.
Let $Q$ be the number of measurements needed to determine this phase in the desired statistical confidence interval, the probability that all the measurements leave the system in its initial state is thus

\begin{eqnarray*}
P_e^{NQ}&\simeq&\lbrack 1-\frac{\tau^2 \Delta H^2_{ee}(1+\cos(\nu))}{2\hbar^2}\rbrack^{\frac{4Q\hbar^2}{\tau^2 \Delta H^2_{ee}\sin(\nu)}}
\end{eqnarray*}

where, supposing condition \ref{condition} to be satisfied, we can neglect the $O(\tau  \bar{\Omega})$ terms .
As $\tau\rightarrow 0$,

\begin{eqnarray*}
P_e^{NQ}&\rightarrow&e^{-\frac{Q}{\text{SNR}}}
\end{eqnarray*}

that can be made arbitrarily close to one by choosing the right $\nu$ and $\tau$.
We have proved that it is indeed possible to measure $ \Delta H_{ee}$ while still impeding
the system evolution.

 \section{Application to actual systems}

With the previous procedure, we can thus measure the mean value of the Hamiltonian's second power  in the initial state. Our final aim would be to determine the squared matrix elements that appear in the Fermi golden rule or, in case of reversible dynamics, the coupling frequencies with the other levels. The possibility to obtain these coefficients from our measurement depends on the details of the system and on the knowledge we have on the structure of the Hamiltonian. There are two simple cases in which we can actually succeed in calculating the transition frequency: first, a two levels system and, secondly, a model in which the initial state is coupled with a continuum of final states and the continuum is narrow enough for the coupling constant not to vary significantly over it (in this last case, we will neglect the possible couplings between the states of the continuum, that do not influence $\Delta H_{ee}$).
In the first case, the Hamiltonian is

\begin{eqnarray}
\label{twolevels}
H = \hbar \left[ \begin{array}{cc} 0 & \Omega  \\ \Omega & \delta  \\
\end{array} \right]
\end{eqnarray}

and $\Delta H_{ee}=\hbar \Omega$, that is we are measuring exactly the Rabi energy of the system. In the second case, the Hamiltonian is

\begin{eqnarray}
H=\sum_i \hbar \omega_i \ket{g_i}\bra{g_i} +\hbar V  \ket{g_i}\bra{e}+\hbar \bar{V} \ket{e}\bra{g_i} \end{eqnarray}

and $\Delta H_{ee}=\hbar \lvert V \lvert \sum_i$, that is we are measuring the desired squared matrix element times the integral of the mode density over the whole continuum.
If we relax the hypothesis of a narrow continuum and we permit the coupling constant to depend on the final state,  we will measure the quantity $\Delta H_{ee}= \hbar\sqrt{\sum_i  \lvert V_i \lvert^2}$, that is the square root of the integral of the squared coupling constant over the continuum. From this quantity, it is not \emph{a priori} possible (unless we know the form of the coupling as a function of the final state up to an overall scaling factor) to determine the quantity that enters the Fermi golden rule. This derives from the fact that, due to the small times implicated, the system actually sees all the final states, not only the resonant ones.

The case of a two levels  system (eq. \ref{twolevels}) has the advantage of allowing simple analytical and numerical calculation, thus giving us the possibility to check our procedure, by performing calculations with the exact value of $P_e$
(i.e. using eq. \ref{P_e_1} instead of eq. \ref{P_e_dev}) and to simulate the exact dynamics of the system and ancilla (i.e. using eq. \ref{tau+} instead of eq. \ref{U_eff_dev}).

\begin{figure}[!]
\begin{center}
\includegraphics[width=8cm]{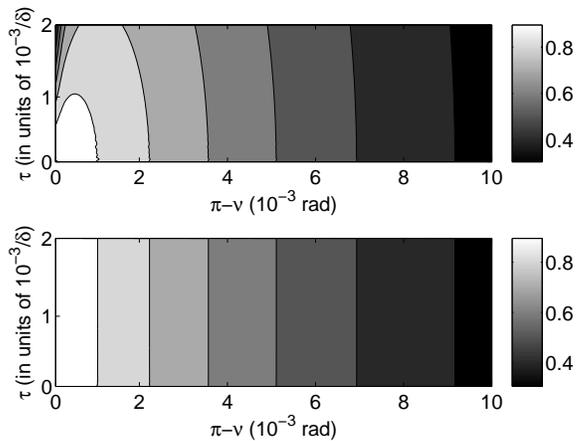}
\caption{\label{due}
The value of $P_e^{NQ}$, with $Q=100$, is plotted using the exact $P_e$ (top panel) and its second order development (bottom panel).}
\end{center}
\end{figure}

In fig. \ref{due}, it is shown a plot of $P_e^{NQ}$  for $\Omega=\delta$, $Q=100$, $N=1/\phi$ with $\phi$ calculated from eq. \ref{phi} and using respectively the exact $P_e$ defined in eq. \ref{P_e_1} (top panel) and its development of eq. \ref{P_e_dev} (bottom panel). In the region where condition \ref{condition} is verified the accord is clear.
Dynamical simulations also show a good accord with the theory. For $\tau=10^{-8}/\delta$ and $\pi-\nu= 10^{-4}$, the simulation gives a measured ratio $\Omega/\delta$ equal to one with a probability $P_e^{NQ}=0.98$, where the values from both panels of fig. \ref{due} are $P_e^{NQ}=0.99$.

\section{Conclusions}


It has been proved that, by carefully entangling a metastable system with an environment,  it is actually possible to measure a certain quantity, linked with the time evolution of the system, while protecting the system from the evolution by means of the quantum Zeno effect and of the entanglement itself.

The measuring procedure here described bears some resemblance to the theory of  protective measurements (\cite{PROT1, PROT2,QP}),  at least for the fact that the wavefunction of the system is protected while measured, and it is thus possible to perform multiple measurements on the same quantum state.

It can be interesting to notice that, being the quantity to be measured the phase $\phi$ of a certain effective evolution operator, the whole measuring procedure can be quite naturally restated in the language of quantum circuits by using a quantum phase estimation algorithm (\cite{NC, KIT}).

\section{Acknowledgements}

I wish to thank C. Ciuti for his help and support during all the stages of this work, G. Ancona, I. Carusotto, M. Chevallier, G. De Chiara and L. De Feo for useful comments and A. Lelli and C. Tauveron for their help in the manuscript preparation.

\end{document}